\documentclass[english,aps,manuscript,preprint]{revtex4-1}
\usepackage[T1]{fontenc}
\usepackage[latin9]{inputenc}
\usepackage{color}
\usepackage{amsmath}
\usepackage{amssymb}
\usepackage{graphicx}
\usepackage{subscript}

\makeatletter

\providecommand{\tabularnewline}{\\}

 \@ifundefined{textcolor}{}
 {%
   \definecolor{BLACK}{gray}{0}
   \definecolor{WHITE}{gray}{1}
   \definecolor{RED}{rgb}{1,0,0}
   \definecolor{GREEN}{rgb}{0,1,0}
   \definecolor{BLUE}{rgb}{0,0,1}
   \definecolor{CYAN}{cmyk}{1,0,0,0}
   \definecolor{MAGENTA}{cmyk}{0,1,0,0}
   \definecolor{YELLOW}{cmyk}{0,0,1,0}
 }

\usepackage{amsthm}
\usepackage{graphics}
\usepackage{times}
\usepackage{bm}
\usepackage{hyperref}
\usepackage{color}
\usepackage{braket}
\renewcommand{\textemdash}{---}

\makeatother

\usepackage{babel}
\begin{document}

\title{Improved semiclassical dynamics through adiabatic switching trajectory
sampling }

\author{Riccardo \surname{Conte}}
\email{riccardo.conte1@unimi.it}

\author{Lorenzo \surname{Parma}}

\author{Chiara \surname{Aieta}}

\author{Alessandro \surname{Rognoni}}

\author{Michele \surname{Ceotto}}
\email{michele.ceotto@unimi.it}

\affiliation{Dipartimento di Chimica, Università degli Studi di Milano, via Golgi
19, 20133 Milano, Italy}
\begin{abstract}
We introduce an improved semiclassical dynamics approach to quantum
vibrational spectroscopy. In this method, a harmonic-based phase space
sampling is preliminarily driven toward non-harmonic quantization
by slowly switching on the actual potential. The new coordinates and
momenta serve as initial conditions for the semiclassical dynamics
calculation, leading to substantial decrease in the number of chaotic
trajectories to deal with. Applications are presented for model and
molecular systems of increasing dimensionality characterized by moderate
or high chaoticity. They include a bidimensional Henon-Heiles potential,
water, formaldehyde, and methane. The method improves accuracy and
precision of semiclassical results and it can be easily interfaced
with all pre-existing semiclassical theories.
\end{abstract}
\maketitle

\section{Introduction}

Chaotic systems can be found in several research fields ranging, for
instance, from physics to meteorology, from chemistry to economy.
They often constitute a hindrance to the possibility of making accurate
predictions and a difficult challenge to face. 

This is also the case for semiclassical (SC) dynamics, which has the
peculiar feature of reproducing quantum effects accurately starting
from classical trajectory runs.\citep{Miller_Atom-Diatom_1970,Heller_SCspectroscopy_1981,Herman_Kluk_SCnonspreading_1984,Kay_Multidim_1994,Grossmann_SConPES_1999,Shalashilin_Child_Coherentstates_2001,Zhang_Pollak_Noprefactor_2004,Miller_PNAScomplexsystems_2005,Zhuang_Ceotto_Hessianapprox_2012,Wehrle_Vanicek_Oligothiophenes_2014,Church_Ananth_Filinov_2017,Buchholz_Ceotto_leakage_2018,Church_Ananth_NonadiabSC_2018}
This hallmark and the possibility to be employed straightforwardly
with any fitted or ``on-the-fly'' potential energy surface (PES)
make SC dynamics appealing for vibrational spectroscopy of complex
molecules\citep{Conte_Ceotto_NH3_2013,Ceotto_watercluster_18,Bertaina_Ceotto_Zundel_2019,Church_Ananth_SCinmixedqcl_2019}
as well as a reference for quantum spectroscopy of medium-large dimensional
systems.\citep{Cina_cheng_2014_variationalquantumclass,Buchholz_Ceotto_MixedSC_2016,Gabas_Ceotto_Glycine_2017,Gabas_Ceotto_SupramolecularGlycines_2018,Kovac_Cina_mixedqscl_2017,Buchholz_Ceotto_applicationMixed_2017,Ceotto_Buchholz_SAM_2018,Patoz_Vanicek_Photoabs_Benzene_2018,Conte_Ceotto_HessianDatabase_2019,Begusic_Vanicek_SingleHessian_2019}\\
The state-of-art is the result of several efforts in the advance of
SC dynamics. A milestone in the development of SC vibrational spectroscopy
is represented by Kaledin and Miller's time-averaged semiclassical
initial value representation (TA SCIVR),\citep{Kaledin_Miller_Timeaveraging_2003,Kaledin_Miller_TAmolecules_2003}
which has permitted to extend applicability of the coherent state
semiclassical Herman Kluk propagator\citep{Kluk_Davis_Highlyexcited_1986}
to small molecules overcoming the well-known convergence issue of
the Monte Carlo phase space integration.\citep{Ceotto_Hase_AcceleratedSC_2013,Tamascelli_Ceotto_GPU_2014,Ma_Ceotto_SN2reactions_2018}\\
Applications to much larger systems are now possible thanks to the
very recent divide-and-conquer semiclassical initial value representation
technique (DC SCIVR), which is based on the projection of the full-dimensional
investigation onto a set of lower dimensional targets.\citep{ceotto_conte_DCSCIVR_2017,DiLiberto_Ceotto_Jacobiano_2018,Gabas_Ceotto_Nucleobasi_2019}
It is useful to remark, though, that for the large systems studied
by means of DC SCIVR a proper Monte Carlo convergence cannot be achieved,
due to the computational overhead that such a computation would require.
Thereby, the simulation must rely on a limited number of trajectories
often evolved ``on-the-fly'' at some accessible \emph{ab initio}
level of electronic structure theory.\\
On this regard, pivotal work by De Leon and Heller has demonstrated
that quantum eigenvalues can be calculated exactly by means of SC
dynamics even employing a single trajectory, provided it has the correct
(unknown) energy.\citep{DeLeon_Heller_SCeigenfunctions_1983} By further
developing this idea, one of us has introduced the multiple coherent
states semiclassical initial value representation (MC SCIVR), whereby
accurate estimates for the quantum frequencies of vibration are obtained
on the basis of a single or handful of trajectories.\citep{Ceotto_AspuruGuzik_Multiplecoherent_2009,Ceotto_AspuruGuzik_PCCPFirstprinciples_2009,Ceotto_AspuruGuzik_Curseofdimensionality_2011,Ceotto_AspuruGuzik_Firstprinciples_2011}\\
These methods restrict the original TA-SCIVR phase space sampling
to a smaller region or even a single point, while the sampling is
done in a harmonic fashion due to the availability of harmonic estimates
at low computational cost even for medium-large molecular systems.
However, the harmonic approximation typically overestimates the true
energy, sometimes even substantially. Furthermore, the actual potential
is not harmonic and the initial harmonic state is not a stationary
state of the molecular Hamiltonian. These aspects contribute to the
numerical instability of the ensuing trajectories and deteriorate
accuracy and precision of semiclassical results.

Adiabatic switching (AS) is a technique that may help overcome these
issues. It has been developed to attain non-harmonic quantization
and sample initial conditions in quasi-classical trajectory (QCT)
simulations. Its foundation lies in the classical adiabatic theorem
which states that action variables are constants of motion during
the evolution of a trajectory lying on a phase-space torus not only
for an isolated system but also in presence of a perturbation, provided
that the latter is switched on very slowly (ideally over an infinite
period of time). \citep{Landau_Lifshitz_1982book,Solovev_AdiabaticInvariants_1978,Skodje_Reinhardt_SCquantization_1985,Johnson_AdiabaticInvariance_1985,Saini_Taylor_AdiabaticSwitching_1988,Huang_Muckerman_AdiabaticSwitching_1995}
AS has also been employed to obtain Wigner distributions\citep{Bose_Makri_WignerAS1_2015,Bose_Makri_WignerAS2_2018}
and for estimates of vibrational energies.\citep{Johnson_H3+AS_1987,Qiyan_Gazdy_AdiabaticSwitching_1988}
Qu and Bowman recently adopted AS to determine the zero-point energy
(ZPE) and fundamental frequencies of a couple of modes of methane,
showing the importance to perform adiabatic switching in an Eckart
frame to get to a narrower and more accurate energy distribution.\citep{Chen_Bowman_AdiabaticSwitching_2016}
Further improvements in precision and accuracy have been later provided
by Nagy and Lendvay by developing AS in internal coordinates to prevent
any kind of ro-vibrational coupling.\citep{Nagy_Lendvay_AdiabaticSwitching_2017}
However, differently from several quantum methods and semiclassical
approaches,\citep{Micciarelli_Ceotto_SCwavefunctions_2018,Micciarelli_Ceotto_SCwavefunctions2_2019}
adiabatic switching is not able to provide eigenfunctions. Furthermore,
AS efficiency is expected to deteriorate for increasing values of
the density of vibrational states, which is known to grow fast with
energy.\citep{Ceotto_Aieta_Gabas_16,Aieta_Ceotto_ParallelSCTST_2019} 

To better point out the focus of this paper, we recall that n-dimensional
integrable systems are those for which n independent integrals of
motion satisfying the Poisson bracket condition can be found and quantization
is doable because the integrals of motion correspond to commuting
observables. In this case trajectories lie on the surface of tori
in phase space. Such trajectories are stable, do not show any chaotic
behavior and never fill up the whole phase space. Conversely, molecular
systems are in general non integrable and trajectories eventually
lead to numerical instability. The basic idea of this work is that
adiabatic switching, starting from the separable and easily quantizable
system made of n harmonic oscillators, can provide an approximate
quantization, which, at least for the short times involved in a semiclassical
spectroscopic calculation, allows use of more stable, quasi-periodic
trajectories. Consequently, the main goal of this manuscript is to
demonstrate that the AS technique allows one to sample the initial
phase space conditions of semiclassical simulations in a more effective
way, decreasing substantially the number of chaotic, numerically unstable
trajectories to deal with, and improving precision and accuracy of
results. We label this ``adiabatically switched'' semiclassical
approach as AS SCIVR. 

In Section \ref{sec:Theoretical-Details} we report on the theoretical
and computational details of the method. Section \ref{sec:Results}
is dedicated to the application of AS SCIVR to a Henon-Heiles model
potential and molecular systems of increasing dimensionality from
water to methane. Finally, we summarize results and discuss possible
future developments and applications of the method in Section \ref{sec:Summary-and-Conclusions}.

\section{Theoretical and Computational Details\label{sec:Theoretical-Details}}

The basic semiclassical working formula we adopted for this paper
is 

\begin{equation}
I(E)=\dfrac{{1}}{(2\pi\hbar)^{N_{vib}}}\int\,d{\bf p}_{0}\int\,d{\bf q}_{0}\dfrac{{1}}{2\pi\hbar T}\left|\int_{0}^{T}\,dt^{\prime}\,e^{i[S_{t^{\prime}}({\bf p}_{0},{\bf q}_{0})+\phi_{t^{\prime}}({\bf p}_{0},{\bf q}_{0})+Et^{\prime}]/\hbar}\langle g_{t^{\prime}}({\bf p}_{0},{\bf q}_{0})|\Psi\rangle\right|^{2}.\label{eq:K-M}
\end{equation}
In Eq. (\ref{eq:K-M}) $I(E)$ is the energy-dependent density of
vibrational states, whose peaks are located at the SC frequencies
of vibration; $N_{vib}$ is the number of vibrational degrees of freedom;
$T$ is the total simulation time; $S_{t^{\prime}}$ is the instantaneous
classical action calculated along the trajectory originated from the
(${\bf p}_{0}$,${\bf q}_{0}$) point in phase space, and $\langle g_{t^{\prime}}({\bf p}_{0},{\bf q}_{0})|\Psi\rangle$
is the quantum mechanical overlap between the coherent state basis
element $|g_{t^{\prime}}({\bf p}_{0},{\bf q}_{0})\rangle$ and the
reference state $|\Psi\rangle$. A coherent state with Gaussian width
matrix $\Gamma$ is defined as

\begin{equation}
\langle{\bf q}|g_{t^{\prime}}({\bf p}_{0},{\bf q}_{0})\rangle=\left(\dfrac{det(\Gamma)}{\pi^{N_{vib}}}\right)^{1/4}\exp\left[-({\bf q}-{\bf q}_{t^{\prime}})^{T}\dfrac{\Gamma}{2}({\bf q}-{\bf q}_{t^{\prime}})+\dfrac{i}{\hbar}{\bf p}_{t^{\prime}}^{T}({\bf q}-{\bf q}_{t^{\prime}})\right],
\end{equation}
where ${\bf p}_{t^{\prime}}$ and ${\bf q}_{t^{\prime}}$ are the
momentum and position vectors at time $t^{\prime}$ obtained upon
classical Hamiltonian evolution from (${\bf p}_{0},{\bf q}_{0}$).
$\Gamma$ is usually chosen to be a diagonal matrix with elements
equal to the harmonic frequencies of vibration. For the calculations
presented here we employed reference states $|\Psi\rangle$ made of
suitable combinations of coherent states centered at equilibrium coordinates
and harmonically estimated momenta, in agreement with our previous
works.\citep{Conte_Ceotto_book_chapter_2019} Finally, $\phi_{t^{\prime}}$
is the phase of the so-called Herman-Kluk prefactor

\begin{equation}
\phi_{t^{\prime}}({\bf p}_{0},{\bf q}_{0})=phase\left[\sqrt{{\left|\dfrac{{1}}{2}\left(\dfrac{\partial{\bf q}_{t^{\prime}}}{\partial{\bf q}_{0}}+\Gamma^{-1}\dfrac{\partial{\bf p}_{t^{\prime}}}{\partial{\bf p}_{0}}\Gamma-i\hbar\dfrac{\partial{\bf q}{}_{t^{\prime}}}{\partial{\bf p}_{0}}\Gamma+\dfrac{i\Gamma^{-1}}{\hbar}\dfrac{\partial{\bf p}{}_{t^{\prime}}}{\partial{\bf q}_{0}}\right)\right|}}\right].\label{eq:HK_prefactor}
\end{equation}
The prefactor is related to deterministic chaos through the monodromy
matrix elements ($\partial{\bf i}/\partial{\bf j}\quad{\bf i}={\bf p}_{t^{\prime}},{\bf q}_{t^{\prime}};\;{\bf j}={\bf p}_{0},{\bf q}_{0}$).
In fact, when one or more of the monodromy matrix eigenvalues start
to grow exponentially in the chaotic regime, numerical integration
of the Herman-Kluk prefactor becomes more and more inaccurate and,
eventually, an unphysical divergence is reached spoiling the entire
SC calculation. Several approaches have been employed to overcome
this issue. The basic one consists in completely discarding trajectories
that reveal a chaotic behavior at some point during the dynamics.
As an alternative, it has been proposed to keep trajectories up to
the instant when numerical instability kicks in, possibly by weighing
their contributions appropriately.\citep{Kay_Numerical_1994,Bertaina_Ceotto_Zundel_2019}
A different way to tackle the problem is by approximating or regularizing
the prefactor.\citep{Guallar_Miller_PrefactorApproximation_1999,Gelabert_Miller_logderivative_2000,DiLiberto_Ceotto_Prefactors_2016,Tatchen_Miller_PrefactorApproximation_2011}
However, none of these approaches is able to provide a way to restrict
the semiclassical calculation to non-chaotic trajectories beforehand. 

\begin{figure}
\begin{centering}
\includegraphics[scale=0.75]{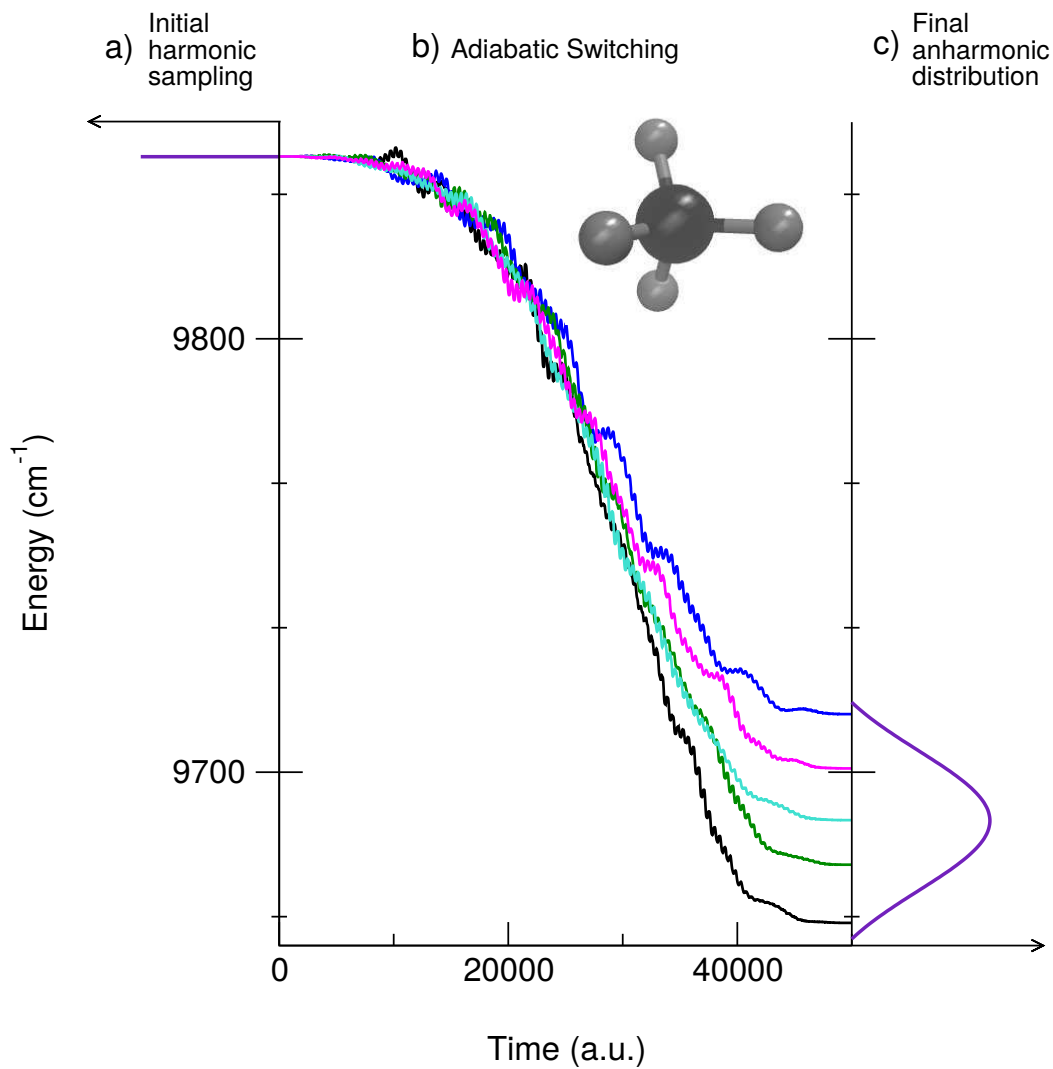}
\par\end{centering}
\caption{Representation of the adiabatic switching procedure for methane. Panel
a): Trajectories are given the harmonic ZPE energy (violet). Panel
b): The energy of 5 trajectories (different colors) is reported as
they evolve under the adiabatic switching Hamiltonian. Panel c): A
final distribution of anharmonic energies (violet) is found. \label{fig:AS_Energy_Distribution_Example}}
\end{figure}

The other technique employed in this work is adiabatic switching.
The AS procedure involves definition of a separable vibrational Hamitonian
($H_{0}$) for which quantization is known or easily achieved, followed
by introduction of the true molecular potential at slow pace until
the fully-coupled vibrational molecular Hamiltonian ($H$) is reached.
In practice, upon calculation of the set of harmonic frequencies of
vibration $\{\omega_{i}\}$, $H_{0}$ is generally chosen to be the
harmonic approximation to $H$ in mass scaled coordinates and momenta 

\begin{equation}
H_{0}=\sum_{i=1}^{N_{vib}}\left(\dfrac{p_{i}^{2}}{2}+\dfrac{\omega_{i}^{2}q_{i}^{2}}{2}\right),\label{eq:H0_normal_modes}
\end{equation}
and the AS Hamiltonian ($H^{AS}$) is a function of time

\begin{equation}
H^{AS}(t)=H_{0}+f_{S}(t)(H-H_{0}).\label{eq:AS_Hamiltonian}
\end{equation}
$f_{S}(t)$ is a switching function selected in agreement with the
literature\citep{Chen_Bowman_AdiabaticSwitching_2016}

\begin{equation}
f_{S}(t)=\dfrac{{t}}{T_{AS}}-\dfrac{{1}}{2\pi}\sin\left(\dfrac{2\pi t}{T_{AS}}\right),\label{eq:Switching_Function}
\end{equation}
which equals 0 at $t=0$ and 1 at $t=T_{AS}$, the total AS simulation
time. For the harmonic Hamiltonian, initial normal mode coordinates
and momenta can be obtained straightforwardly from action-angle variables,
i.e. $q_{i}=[(2n_{i}+1)\hbar/\omega_{i}]^{1/2}\cos\zeta_{i};\quad p_{i}=-[(2n_{i}+1)\hbar\omega_{i}]^{1/2}sin\,\zeta_{i}$.
$n_{i}$ are integer actions, while $\zeta_{i}$ are randomly selected
angles from a uniform distribution. Classical dynamics is then performed
for a time $T_{AS}$ under the Hamiltonian $H^{AS}(t)$. Clearly,
during adiabatic switching, the total energy is not conserved. It
starts from the harmonic value and ends at an estimate of the energy
of the corresponding quantized state of the actual molecular Hamiltonian.
From an ensemble of AS trajectories, one eventually gets a distribution
that approximates the energy of the state, as shown for methane in
Figure \ref{fig:AS_Energy_Distribution_Example}. We employed a pre-existing
methane PES by Lee, Martin and Taylor.\citep{lee_taylor_PESch4_1995}
$T_{AS}$ was chosen equal to 1.21 ps (50000 atomic units), a time
step of 0.242 fs was employed, and the dynamics, as for all other
investigations presented in this paper, was integrated by means of
a 4-th order symplectic algorithm with a fixed step equal to 10\textsuperscript{-3}
for finite difference calculations.\citep{Brewer_Manolopoulos_15dof_1997} 

\begin{figure}
\begin{centering}
\includegraphics[scale=0.6]{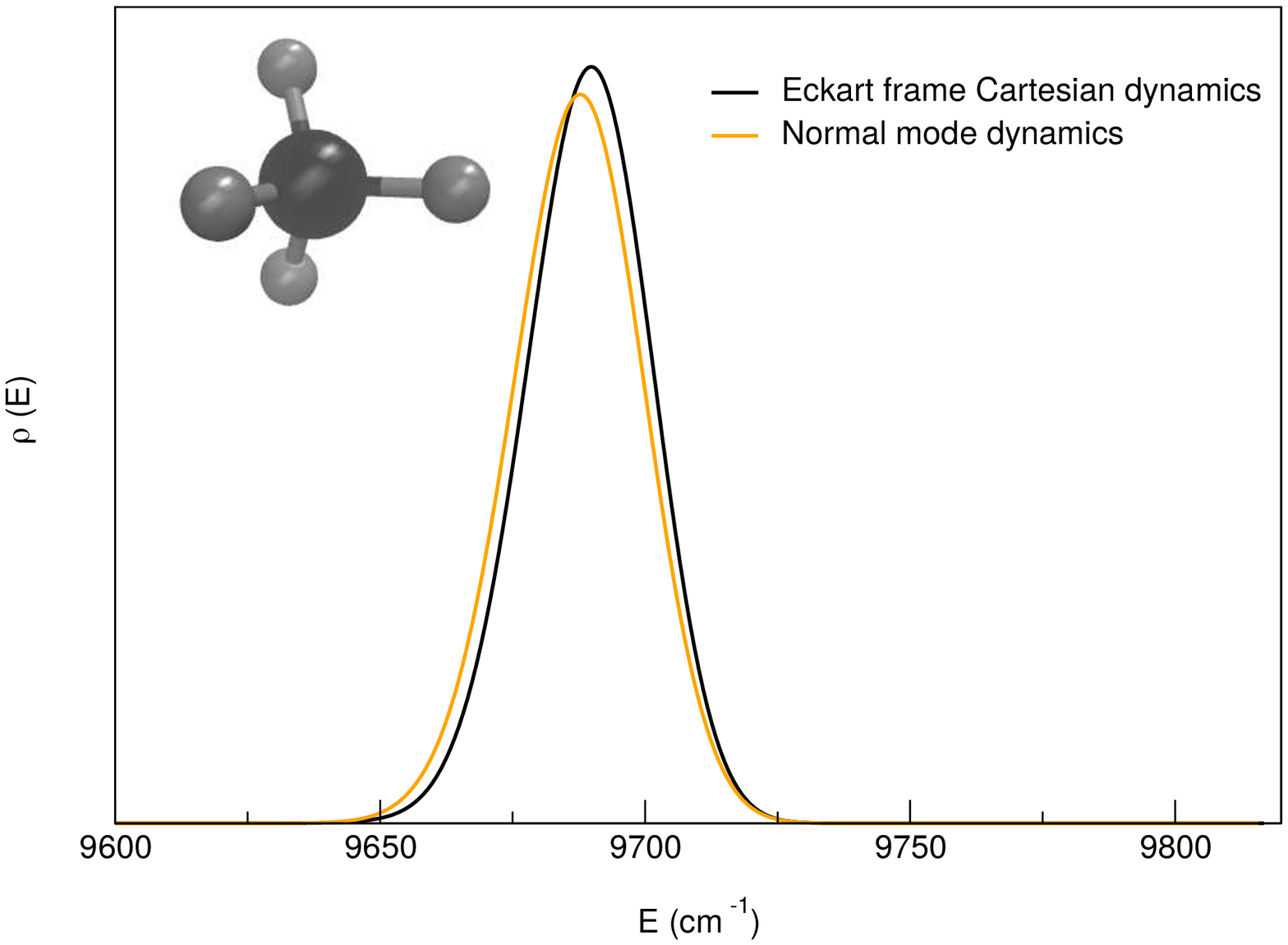}
\par\end{centering}
\caption{Final AS energy distributions for methane started with harmonic zero
point energy (9842 cm\protect\textsuperscript{-1}). Comparison is
between Cartesian dynamics in Eckart frame (maroon), as described
in Ref. \citenum{Chen_Bowman_AdiabaticSwitching_2016}, and the normal
mode dynamics employed in this work (orange).\label{fig:Final-AS-distributions} }
\end{figure}

Because adiabatic switching is known to work more efficiently at low
density of vibrational states and for not strongly coupled systems,\citep{Qiyan_Gazdy_AdiabaticSwitching_1988}
we employed it to get an initial distribution in phase space for our
subsequent and more widely applicable semiclassical dynamics simulations.
In other words, the outcome of the adiabatic switching procedure served
as an initial sampling for the SCIVR spectral calculations. We evolved
the dynamics in normal modes in agreement with our past standard TA-SCIVR
applications. Fig. \ref{fig:Final-AS-distributions} shows a comparison
for methane between the AS final energy distributions of 9000 trajectories
obtained starting from harmonic ZPE sampling by means of the approach
reported in Ref. \citenum{Chen_Bowman_AdiabaticSwitching_2016} and
our normal-mode based one. Computational details are the same as previously
reported. Results are in strict agreement. We removed the ro-vibrational
coupling in our normal mode reference frame by not evolving the rotational
degrees of freedom, an artefact which, on the other hand, slightly
perturbs the total angular momentum, owing to the loss of reliability
of normal modes out of equilibrium. This led to the very small (but
negligible for our purposes) discrepancy between the two simulations.
For the Gaussian envelop of bins a width of 7.2 cm\textsuperscript{-1}
has been adopted in all simulations.

As for TA SCIVR and the semiclassical part of our AS-SCIVR simulations,
to determine whether a classical trajectory had to be discarded or
not, we compared along the dynamics the shift from unity of the monodromy
matrix determinant to an arbitrary threshold $\sigma$.\citep{Wang_Miller_GeneralizedFilinov_2001,Kaledin_Miller_Timeaveraging_2003}
Whenever the shift was larger than the chosen $\sigma$ we eliminated
the whole trajectory from the set of those contributing to the final
spectrum.

\section{Results\label{sec:Results}}

To demonstrate the performance of AS SCIVR we applied it to a set
of systems, characterized by different regimes of trajectory rejection,
and compared the outcomes with the corresponding ones obtained by
using a standard TA-SCIVR procedure. Results were also tested against
available quantum mechanical benchmarks.

\subsection{Henon-Heiles model}

\begin{figure}
\begin{centering}
\includegraphics[scale=0.55]{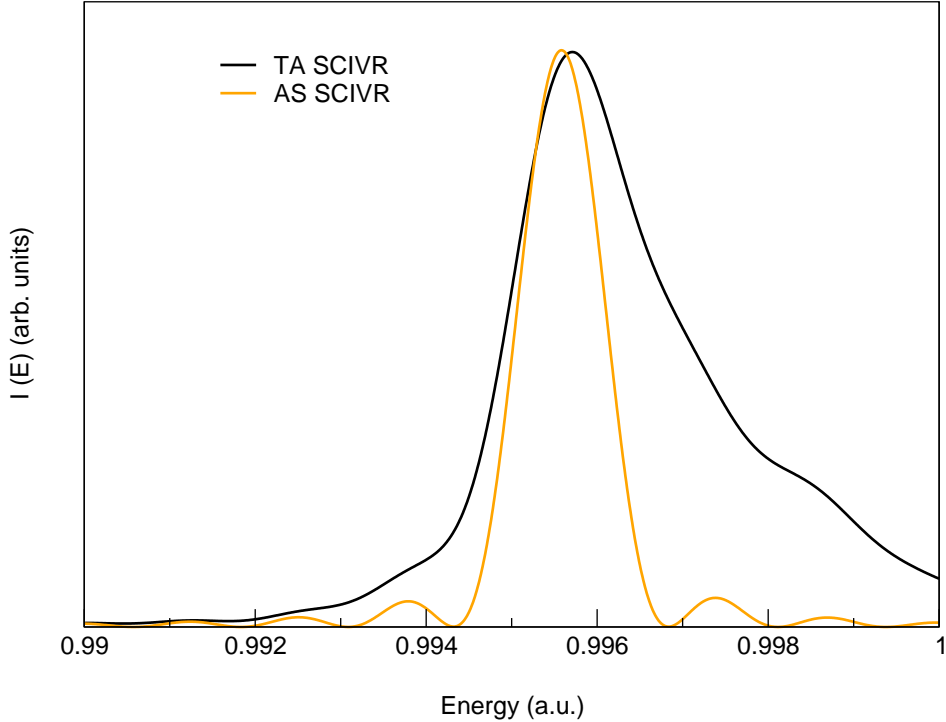}
\par\end{centering}
\caption{Detail of the ZPE signal for the investigated Henon-Heiles system,
as obtained from a standard TA-SCIVR approach (black) and AS SCIVR
(orange). Intensities have been scaled to get matching maximum values.\label{fig:ZPE-signal-HH}}
\end{figure}

We start presenting an application to a low-dimensional model system
characterized by moderate chaos. We chose the following two-dimensional
Henon-Heiles potential, which was employed also in previous works\citep{Huber_Heller_Henon-Heiler-Kay_1988,Huber_Heller_HybridDynamics_1989,Kay_Multidim_1994}

\begin{equation}
V(q_{1},q_{2})=\dfrac{1}{2}\omega_{1}^{2}q_{1}^{2}+\dfrac{1}{2}\omega_{2}^{2}q_{2}^{2}+\lambda q_{2}(q_{1}^{2}+\eta q_{2}^{2})\quad\omega_{1}=1.3,\,\omega_{2}=0.7,\,\lambda=-0.1,\,\eta=0.1.\label{eq:Hen-Heiles}
\end{equation}
Values of the parameters in Eq. (\ref{eq:Hen-Heiles}) are given in
atomic units (a.u.). This leads to a different time scale for the
dynamics with respect to the case of methane. In particular, $T_{AS}$
was set to 12.1 fs and $T$ was selected equal to about 121 fs with
a timestep of 0.00242 fs. A different SC calculation for each of the
first 8 eigenvalues was performed by means of both AS SCIVR and TA
SCIVR. Initial conditions were determined either by centering a Husimi
distribution at the harmonic energy of the target eigenvalue (TA-SCIVR
simulations), or by starting a preliminary AS procedure from the relevant
harmonic quantization (AS-SCIVR simulations). TA SCIVR featured a
trajectory rejection rate ranging from about 36\% to 79\% given a
threshold $\sigma=10^{-6}$. Under the same strict condition, all
trajectories generated for AS-SCIVR simulations were instead suitable
to be employed. Table \ref{tab:Eigenvalues_Hen-Heiles} shows a comparison
of the first 8 eigenvalues obtained by means of the discrete variable
representation method (DVR), TA SCIVR, and AS SCIVR. For the sinc-DVR
calculation\citep{colbert_miller_dvr_1992} we employed a rectangular
grid ({[}-5:5{]}, {[}-8:8{]}) with 70 points per each dimension without
any energy cutoff.
\begin{table}
\centering{}\caption{Calculated eigenvalues and full widths at half maximum (in parenthesis)
for the first 8 energy levels of a 2-dimensional Henon Heiles model.
Values are in atomic units. Under the column for Level the corresponding
harmonic excitation is given in parenthesis. \label{tab:Eigenvalues_Hen-Heiles}}
\begin{tabular}{|c|c|c|c|}
\hline 
Level & DVR & TA SCIVR & AS SCIVR\tabularnewline
\hline 
\hline 
0 (ZPE) & 0.996 & 0.996 (0.002) & 0.996 (0.001)\tabularnewline
\hline 
1 ($\omega_{2}$) & 1.687 & 1.687 (0.003) & 1.687 (0.001)\tabularnewline
\hline 
2 ($\omega_{1}$) & 2.278 & 2.278 (0.003) & 2.278 (0.001)\tabularnewline
\hline 
3 (2$\omega_{2}$) & 2.375 & 2.375 (0.003) & 2.375 (0.001)\tabularnewline
\hline 
4 ($\omega_{1}+\omega_{2}$) & 2.958 & 2.959 (0.003) & 2.958 (0.002)\tabularnewline
\hline 
5 (3$\omega_{2}$) & 3.060 & 3.060 (0.004) & 3.060 (0.002)\tabularnewline
\hline 
6 (2$\omega_{1}$) & 3.548 & 3.548 (0.005) & 3.548 (0.001)\tabularnewline
\hline 
7 (2$\omega_{2}+\omega_{1}$) & 3.635 & 3.635 (0.008) & 3.635 (0.001)\tabularnewline
\hline 
\end{tabular}
\end{table}
 Both semiclassical simulations provide results in perfect agreement
with the DVR benchmark, spanning overtones and combined excitations,
but TA SCIVR yields somewhat less precise estimates. This is related
to the different widths of the spectral features obtained by means
of the two SC approaches. AS SCIVR indeed returns not only accurate
but also very precise results due to the small full-width at half
maximum (FWHM) values of its signals. FWHM data are definitely larger
for TA SCIVR. The better quality of the AS-SCIVR signals is also demonstrated
by the fact that well-defined, narrow peaks can be obtained for all
8 eigenvalues employing just the reference state centered at the harmonic
ZPE energy. In the case of TA SCIVR if the reference state is not
tailored on the state under investigation, then extended bands with
several peaks rather than single signals are eventually found for
levels 6 and 7. Figure \ref{fig:ZPE-signal-HH} allows to fully appreciate
the increased precision of an AS-SCIVR calculation in evaluating the
ZPE. In fact, while a very well resolved signal is found for the AS-SCIVR
simulation, in the case of TA SCIVR a much larger and asymmetric peak
is recovered.

\subsection{H\protect\textsubscript{2}O}

\begin{figure}
\begin{centering}
\includegraphics[scale=0.6]{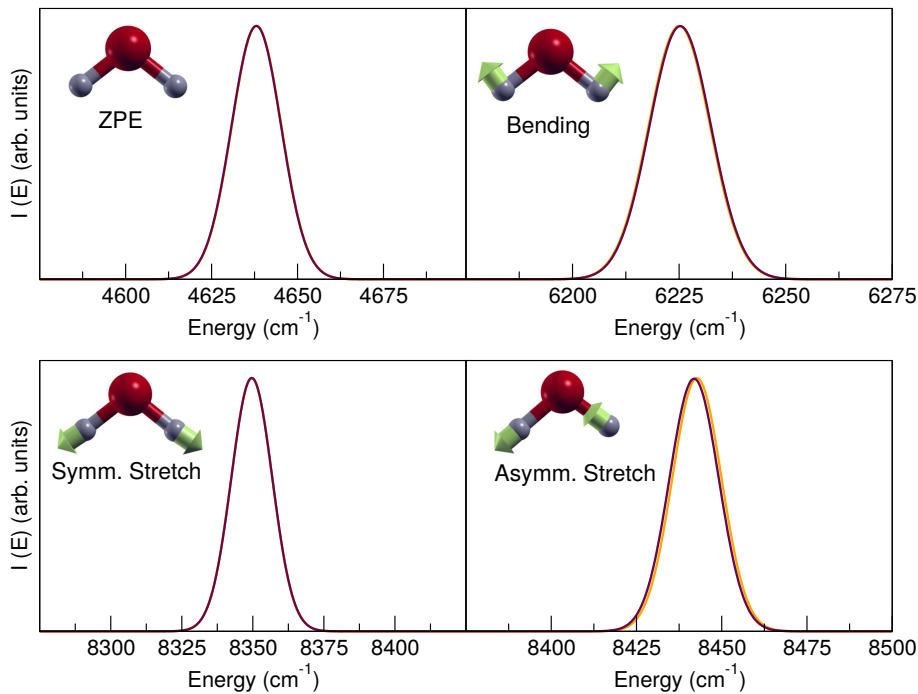}
\par\end{centering}
\caption{Comparison of adiabatic switching energy distributions for H\protect\textsubscript{2}O
obtained with Eckart frame Cartesian dynamics (maroon) and normal
mode dynamics (orange). The initial harmonic energies are equal to
ZPE in panel a); bending excitation in panel b); symmetric stretch
excitation in panel c); asymmetric stretch excitation in panel d).
The width of the Gaussian envelop of bins was chosen equal to 7.2
cm\protect\textsuperscript{-1}.\label{fig:AS_H2O}}
\end{figure}

Water is the first molecule we studied. It is characterized by 3 vibrational
degrees of freedom and the well-known Fermi resonance involving the
bending overtone and the symmetric stretch. To start with the calculations,
we generated AS energy distributions for the ZPE and the energy levels
corresponding to the first excitation of the three vibrational modes.
Figure \ref{fig:AS_H2O} shows the similarity of the distributions
obtained using either normal mode dynamics or Cartesian dynamics in
Eckart frame. We employed the analytical surface by Dressler and Thiel\citep{Dressler_Thiel_WaterPES_1997}.
For the AS procedure we adopted a time step of 10 a.u. for a total
$T_{AS}$ time of about 1.2 ps. 

For the semiclassical simulations the same time step and total simulation
time $T$ were employed. For each AS-SCIVR simulation a distribution
of 3 000 initial conditions was obtained upon performing adiabatic
switching starting from the harmonic quantization corresponding to
the target state, while in the case of the TA-SCIVR calculation a
Husimi distribution of 3 000 initial conditions centered at the harmonic
ZPE was employed. First, we looked at the fraction of trajectories
to be discarded for values of $\sigma$ ranging from 10\textsuperscript{-2}
to 10\textsuperscript{-6}. The TA-SCIVR simulation returned percentages
of rejection between 10.1\% and 65.4\%, while, remarkably, AS SCIVR
could rely on the entire set of trajectories independently of the
$\sigma$ threshold.\\
Moving to the frequencies of vibration, Table \ref{tab:freq_H2O}
compares the quantum mechanical results obtained by means of a Lanczos
algorithm and reported in the Supplementary Information of Ref. \citenum{Micciarelli_Ceotto_SCwavefunctions_2018}
to the outcomes of TA SCIVR and the new AS-SCIVR technique. $\sigma$
was set equal to 10\textsuperscript{-2}, a typical figure we adopt
in molecular calculations. Results are slightly better for the AS-SCIVR
calculation, whose signals are much more precise as clearly pointed
out by the lower FWHM values. However, we notice that most of the
error is due to the ZPE estimate. We will discuss more on this point
in the final Section of the paper.

\begin{figure}
\begin{centering}
\includegraphics[scale=0.55]{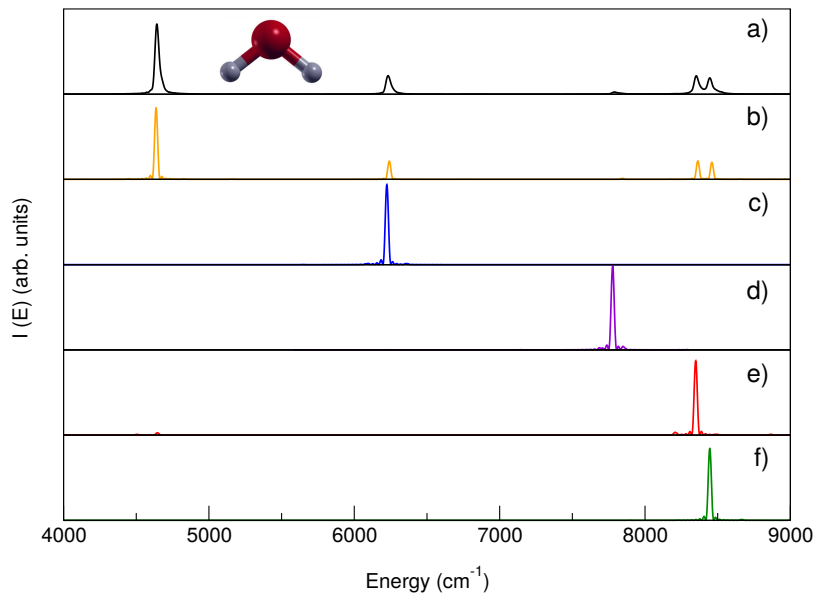}
\par\end{centering}
\caption{Power spectra of H\protect\textsubscript{2}O. Panel a): TA-SCIVR
simulation; Panel b): AS-SCIVR simulation from ZPE AS distribution;
Panel c) - f): AS-SCIVR simulations from bending, bending overtone,
symmetric stretch and asymmetric stretch AS distributions, respectively.\label{fig:Power-spectra_H2O} }
\end{figure}

\begin{table}
\begin{centering}
\caption{ZPE and first vibrational frequencies of H\protect\textsubscript{2}O.
Frequencies associated to Levels 1-4 are obtained by difference between
the corresponding energy level and the ZPE value. Under the Level
or Frequency column the harmonic excitation label is given in parenthesis
( $\omega_{b}$ for the bending; $\omega_{s}$ for the symmetric stretch;
$\omega_{a}$ for the asymmetric stretch). Under the TA SCIVR and
AS SCIVR columns, FWHM values are given in parentheses. QM indicates
the quantum mechanical benchmark; label HARM is the column of harmonic
estimates; MAE stands for mean absolute error. All values are in cm\protect\textsuperscript{-1}.\label{tab:freq_H2O}}
\par\end{centering}
\centering{}%
\begin{tabular}{|c|c|c|c|c|}
\hline 
Level or Frequency & QM\textsuperscript{\citep{Micciarelli_Ceotto_SCwavefunctions_2018}} & TA SCIVR & AS SCIVR & HARM\tabularnewline
\hline 
\hline 
1 ($\omega_{b}$) & 1587 & 1590 (42) & 1587 (24) & 1650\tabularnewline
\hline 
2 (2 $\omega_{b}$) & 3139 & 3147 (60) & 3140 (24) & 3300\tabularnewline
\hline 
3 ($\omega_{s}$) & 3716 & 3711 (41) & 3713 (24) & 3831\tabularnewline
\hline 
4 ($\omega_{a}$) & 3803 & 3804 (41) & 3808 (24) & 3941\tabularnewline
\hline 
ZPE & 4660 & 4642 (34) & 4637 (24) & 4711\tabularnewline
\hline 
MAE & - & 7 & 6 & 105\tabularnewline
\hline 
\end{tabular}
\end{table}
Figure \ref{fig:Power-spectra_H2O} presents the power spectra. In
particular, from a comparison between the plots reporting the complete
spectrum and based on ZPE distributions, it is clear that AS SCIVR
gives more precise estimates (this is most evident looking at the
symmetric and asymmetric stretches). On the other hand, an AS-SCIVR
simulation started from harmonic ZPE quantization yields a harmonic
estimate for the overtone, which needs a tailored simulation to be
detected correctly.

\subsection{H\protect\textsubscript{2}CO}

The second molecular system we studied was formaldehyde. We used a
pre-existing PES by Martin, Lee, and Taylor.\citep{martin_taylo_PESch2o_1993}
Similarly to the water investigation, for AS we employed a time step
of 10 a.u. and a total time $T_{AS}$ of about 1.2 ps. The same values
were adopted for TA-SCIVR calculations and the semiclassical part
of AS-SCIVR simulations. In all instances a total of 6000 trajectories
was run. We do not report AS energy distribution plots for H\textsubscript{2}CO
but, once more, there is utmost agreement between the normal mode
and Cartesian approaches. The threshold for trajectory rejection was
set to 10\textsuperscript{-2}. Rejection percentages are reported
in Table \ref{tab:Rej_H2CO}, and we notice that, also in this case,
AS SCIVR helps a lot in reducing substantially the fraction of discarded
trajectories. 

\begin{table}
\caption{Percentage of trajectory rejection in TA-SCIVR and AS-SCIVR simulations
of H\protect\textsubscript{2}CO ($T$ $\approx$ 0.60 ps and $T$
$\approx$ 1.21 ps) for several rejection thresholds.\label{tab:Rej_H2CO}}

\centering{}%
\begin{tabular}{|c||c|c||c|c|}
\cline{2-5} 
\multicolumn{1}{c|}{} & \multicolumn{2}{c||}{$T$ $\approx$ 0.60 ps} & \multicolumn{2}{c|}{$T$ $\approx$ 1.21 ps}\tabularnewline
\hline 
$\sigma$ & TA SCIVR & AS SCIVR & TA SCIVR & AS SCIVR\tabularnewline
\hline 
\hline 
10\textsuperscript{-2} & 47.6\% & 0.2\% & 83.5\% & 22.3\%\tabularnewline
\hline 
10\textsuperscript{-3} & 56.7\% & 0.9\% & 87.2\% & 38.2\%\tabularnewline
\hline 
10\textsuperscript{-4} & 65.9\% & 3.0\% & 90.9\% & 57.8\%\tabularnewline
\hline 
10\textsuperscript{-5} & 75.3\% & 8.6\% & 94.0\% & 77.3\%\tabularnewline
\hline 
10\textsuperscript{-6} & 84.2\% & 25.2\% & 97.0\% & 93.1\%\tabularnewline
\hline 
\end{tabular}
\end{table}
Moving to the analysis of the frequencies of vibration, we first focus
on fundamentals only. For water and the Henon-Heiles model we performed
specific AS-SCIVR calculations for each spectral feature. In the case
of H\textsubscript{2}CO we wanted to assess the accuracy of a single
AS-SCIVR simulation started from harmonic ZPE quantization in estimating
the fundamental frequencies. Table \ref{tab:Fundamental-frequencies-of-H2CO}
demonstrates that the numerical outcome is very similar to the TA-SCIVR
one.

\begin{figure}
\begin{centering}
\includegraphics[scale=0.65]{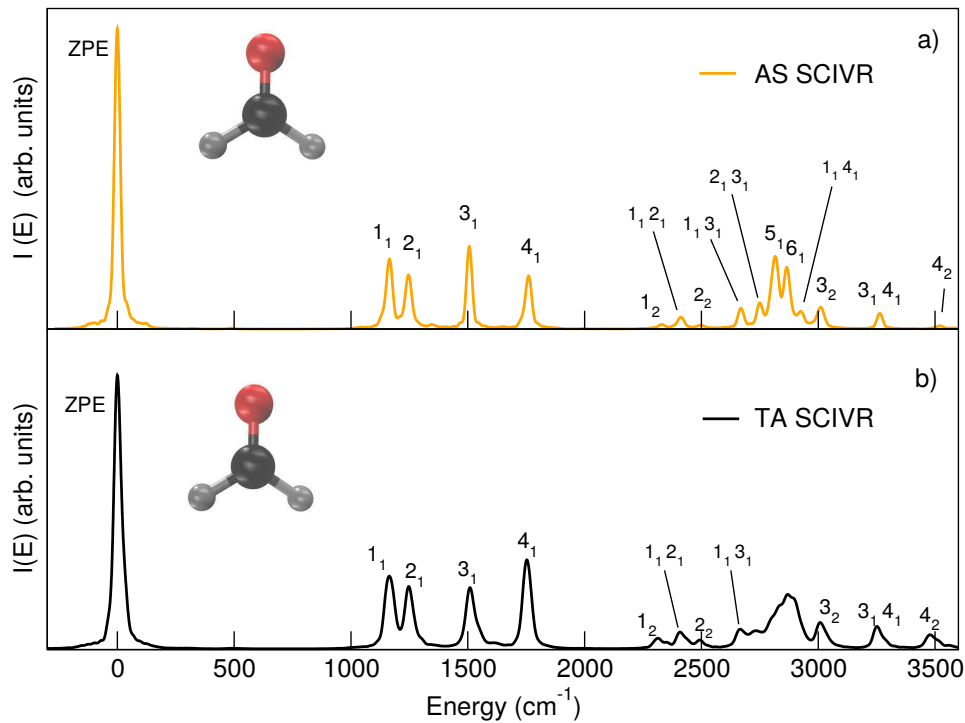}
\par\end{centering}
\caption{Comparison between AS-SCIVR (panel a), orange) and TA-SCIVR (panel
b), black) power spectra of formaldehyde.\label{fig:H2CO_spectra}}
\end{figure}

\begin{table}
\centering{}\caption{Fundamental frequencies of vibration for H\protect\textsubscript{2}CO
from TA-SCIVR and AS-SCIVR simulations based on the harmonic ZPE.
Under the Frequency column, the harmonic excitation label is given.
QM indicates the quantum mechanical benchmark obtained through a variational
approach; label HARM is for the column of harmonic estimates; MAE
stands for mean absolute error. FWHM values are given in parentheses.
N/A points out that a FWHM value could not be determined. All values
are in cm\protect\textsuperscript{-1}.\label{tab:Fundamental-frequencies-of-H2CO} }
\begin{tabular}{|c|c|c|c|c|}
\hline 
Frequency & QM\textsuperscript{\citep{carter_handy_exactCH2O_1995}} & TA SCIVR & AS SCIVR & HARM\tabularnewline
\hline 
\hline 
$\omega_{1}$ & 1171 & 1164 (52) & 1165 (34) & 1192\tabularnewline
\hline 
$\omega_{2}$ & 1253 & 1247 (46) & 1247 (34) & 1275\tabularnewline
\hline 
$\omega_{3}$ & 1509 & 1509 (48) & 1507 (28) & 1543\tabularnewline
\hline 
$\omega_{4}$ & 1750 & 1753 (45) & 1760 (31) & 1781\tabularnewline
\hline 
$\omega_{5}$ & 2783 & 2810 (N/A) & 2816 (43) & 2929\tabularnewline
\hline 
$\omega_{6}$ & 2842 & 2879 (N/A) & 2865 (42) & 2996\tabularnewline
\hline 
MAE & - & 13 & 13 & 68\tabularnewline
\hline 
\end{tabular}
\end{table}
Differences can be spotted by looking at the corresponding power spectra.
Figure \ref{fig:H2CO_spectra} reports them. It is clear that the
AS-SCIVR procedure provides a better resolution of the spectral signals
and helps with the assignment. This is most evident for the band involving
the fifth and sixth fundamentals, which TA SCIVR is not able to identify
adequately. For this reason, the TA-SCIVR values of $\omega_{5}$
and $\omega_{6}$ in Table \ref{tab:Fundamental-frequencies-of-H2CO}
are just tentative and driven by knowledge of the quantum mechanical
values. To improve the quality of results, at this point the standard
TA-SCIVR procedure requires additional runs with tailored reference
states, but, if more than a single simulation is allowed, then targeted
AS-SCIVR simulations are able to provide more accurate and, most of
all, precise estimates, as reported in Table \ref{tab:H2CO_complete}.
For these refined calculations we employed tailored reference states
to separate $\omega_{5}$ and $\omega_{6}$ in TA-SCIVR simulations,
while we performed 6 different calculations, each one started with
one quantum of harmonic excitation in one of the 6 modes, for the
AS-SCIVR case. Tailored TA SCIVR could resolve between $\omega_{5}$
and $\omega_{6}$, but at the cost of very large peak amplitudes.
For refined AS SCIVR the MAE with respect to the quantum mechanical
benchmark, computed on the first 16 frequencies, is down to 8 cm\textsuperscript{-1}. 

\begin{table}
\centering{}\caption{TA-SCIVR and AS-SCIVR estimates for the first 16 frequencies of vibration
for H\protect\textsubscript{2}CO. Under the Frequency column, the
harmonic excitation label is given. QM indicates the quantum mechanical
benchmark; label HARM is for the column of harmonic estimates; MAE
stands for mean absolute error. N/A points out that a FWHM value could
not be determined. Values are in cm\protect\textsuperscript{-1}.
\label{tab:H2CO_complete}}
\begin{tabular}{|c|c|c|c|c||c|c|c|c|c|}
\hline 
Frequency & QM\textsuperscript{\citep{carter_handy_exactCH2O_1995}} & TA SCIVR & AS SCIVR & HARM & Frequency & QM\textsuperscript{\citep{carter_handy_exactCH2O_1995}} & TA SCIVR & AS SCIVR & HARM\tabularnewline
\hline 
\hline 
$\omega_{1}$ & 1171 & 1164 (52) & 1158 (29) & 1192 & $\omega_{2}+\omega_{3}$ & 2729 & 2732 (N/A) & 2724 (35) & 2818\tabularnewline
\hline 
$\omega_{2}$ & 1253 & 1247 (46) & 1245 (30) & 1275 & $\omega_{5}$ & 2783 & 2813 (98) & 2784 (49) & 2929\tabularnewline
\hline 
$\omega_{3}$ & 1509 & 1509 (48) & 1507 (29) & 1543 & $\omega_{6}$ & 2842 & 2861 (85) & 2844 (33) & 2996\tabularnewline
\hline 
$\omega_{4}$ & 1750 & 1753 (45) & 1748 (29) & 1781 & $\omega_{1}+\omega_{4}$ & 2913 & 2893 (N/A) & 2908 (36) & 2973\tabularnewline
\hline 
2$\omega_{1}$ & 2333 & 2313 (78) & 2315 (32) & 2384 & $\omega_{2}+\omega_{4}$ & 3007 & 3007 (56) & 3009 (37) & 3056\tabularnewline
\hline 
$\omega_{1}+\omega_{2}$ & 2431 & 2408 (58) & 2406 (30) & 2467 & 2$\omega_{3}$ & 3016 & 3007 (56) & 3016 (28) & 3086\tabularnewline
\hline 
2$\omega_{2}$ & 2502 & 2492 (N/A) & 2490 (30)  & 2550 & $\omega_{3}+\omega_{4}$ & 3250 & 3252 (48) & 3261 (30) & 3324\tabularnewline
\hline 
$\omega_{1}+\omega_{3}$ & 2680 & 2667 (N/A) & 2664 (33) & 2735 & 2$\omega_{4}$ & 3480 & 3478 (67) & 3488 (31) & 3562\tabularnewline
\hline 
\multicolumn{1}{c}{} & \multicolumn{1}{c}{} & \multicolumn{1}{c}{} & \multicolumn{1}{c}{} & \multicolumn{1}{c|}{} & MAE & - & 10 & 8 & 64\tabularnewline
\cline{6-10} 
\end{tabular}
\end{table}

\subsection{CH\protect\textsubscript{4}}

The final molecule we present is methane, whose PES and AS energy
distribution obtained starting from harmonic ZPE quantization have
already been illustrated (see Figures \ref{fig:AS_Energy_Distribution_Example}
and \ref{fig:Final-AS-distributions}). For this system we decided
to perform a single simulation with both AS SCIVR and TA SCIVR including
all fundamentals, an overtone, and a combined excitation. This allows
us to point out the advantages of AS SCIVR over TA SCIVR directly. 

\begin{figure}
\begin{centering}
\includegraphics[scale=0.65]{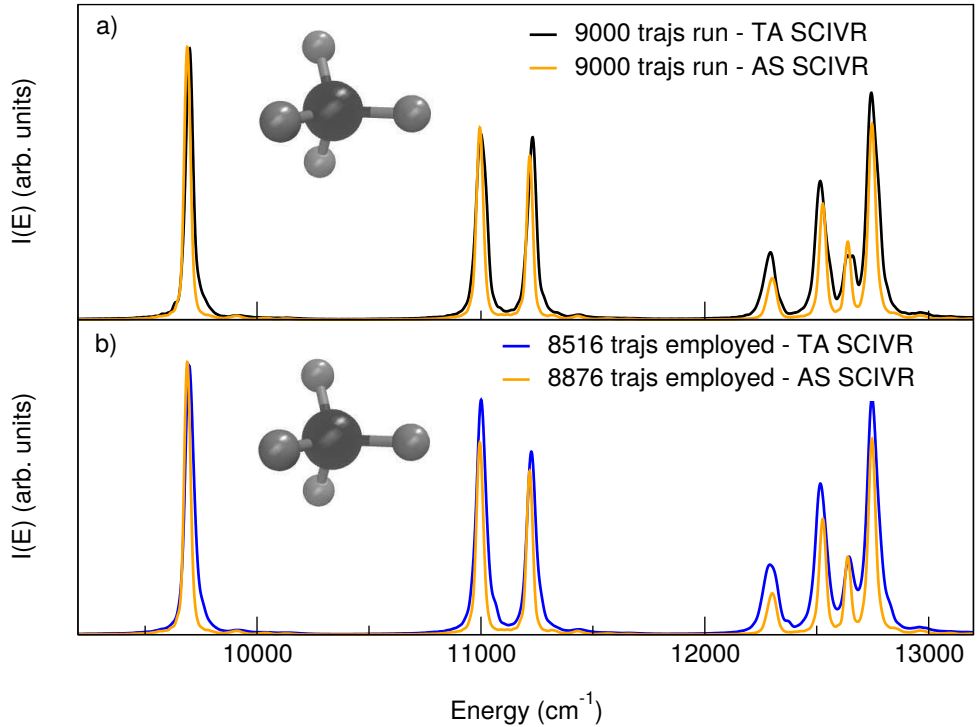}
\par\end{centering}
\caption{Comparison between methane power spectra obtained from AS-SCIVR (orange)
and TA-SCIVR (black/blue) calculations. Panel a): 9 000 trajectories
run in both cases. Panel b): Simulations based on similar numbers
of non-discarded trajectories. Intensities have been scaled to get
matching ZPEs. \label{fig:methane_spectra}}
\end{figure}
Figure \ref{fig:methane_spectra} is made of two comparisons between
AS-SCIVR results obtained starting from harmonic ZPE quantization,
and standard TA-SCIVR outcomes collected from a Husimi distribution
of initial conditions centered around the harmonic ZPE. We adopted
a timestep of 0.242 fs with $T=1.21$ ps, and a rejection threshold
$\sigma=10^{-2}$. In the first case both simulations were based on
9000 trajectories. While 8876 ($\approx$ 98.6\%) of those employed
in AS SCIVR were retained for the SC calculation, only 897 trajectories
started from the Husimi distribution were kept ($\approx$ 10\%).
In the second case, we increased to 90 000 the number of trajectories
for the TA-SCIVR calculation. In this way 8 516 trajectories were
retained to build the TA-SCIVR spectrum, a number comparable to the
AS-SCIVR instance. It is clear from Figure \ref{fig:methane_spectra}
that AS SCIVR provides much narrower and more precise signals. However,
this is not only due to the higher number of trajectories retained
to build the AS-SCIVR spectrum, as the first panel of Figure \ref{fig:methane_spectra}
might suggest, but it is a true hallmark of the method as confirmed
by the bottom panel of the same Figure, where the number of trajectories
contributing to the spectrum is comparable.
\begin{table}
\centering{}\caption{Percentage of trajectory rejection in TA-SCIVR and AS-SCIVR simulations
of CH\protect\textsubscript{4} ($T\approx$ 0.60 ps and $T\approx$
1.21ps) for several rejection thresholds.\label{tab:Rej_Percentage}}
\begin{tabular}{|c||c|c||c|c|}
\cline{2-5} 
\multicolumn{1}{c|}{} & \multicolumn{2}{c||}{$T$ = 0.60 ps} & \multicolumn{2}{c|}{$T$ = 1.21 ps}\tabularnewline
\hline 
$\sigma$ & TA SCIVR & AS SCIVR & TA SCIVR & AS SCIVR\tabularnewline
\hline 
\hline 
10\textsuperscript{-2} & 54.6\% & 0.0\% & 90.4\% & 1.4\%\tabularnewline
\hline 
10\textsuperscript{-3} & 61.7\% & 0.0\% & 94.0\% & 5.3\%\tabularnewline
\hline 
10\textsuperscript{-4} & 70.7\% & 0.0\% & 97.3\% & 16.6\%\tabularnewline
\hline 
10\textsuperscript{-5} & 81.1\% & 0.1\% & 99.0\% & 42.2\%\tabularnewline
\hline 
10\textsuperscript{-6} & 90.2\% & 1.1\% & 99.7\% & 79.4\%\tabularnewline
\hline 
\end{tabular}
\end{table}
\\
Table \ref{tab:Rej_Percentage} demonstrates even further the importance
of AS SCIVR compared to TA SCIVR in decreasing the number of trajectories
displaying a chaotic behavior. As expected, it is also possible to
appreciate that numerical stability is worse conserved for higher
values of T. Nevertheless, we were able to perform our AS-SCIVR simulations
of methane with virtually no trajectory rejection (1.4\%). In addition
to being more precise, we notice that signals coming from the AS-SCIVR
simulation are more accurate when compared to available quantum mechanical
results. Table \ref{tab:CH4_Freq_Rej} points out these aspects, reporting
that for AS SCIVR the mean absolute error is down to just 7 wavenumbers
with respect to the quantum mechanical benchmark.
\begin{table}
\centering{}\caption{Unique frequency values of methane from TA-SCIVR and AS-SCIVR simulations
based on a similar number of retained trajectories. Under the Level
or Frequency column the harmonic excitation label is given. QM indicates
the quantum mechanical benchmark \textcolor{black}{obtained with vibrational
self-consistent field theory and a variational approach}; label HARM
is for the column of harmonic estimates; MAE stands for mean absolute
error. FWHM data are reported in parentheses. N/A points out that
a FWHM value could not be determined. All values are in cm$^{\text{-1}}$.
\label{tab:CH4_Freq_Rej}}
\begin{tabular}{|c|c|c|c|c|}
\hline 
Level or Frequency & QM\textsuperscript{\citep{Carter_Bowman_Methane_1999}} & TA SCIVR & AS SCIVR & HARM\tabularnewline
\hline 
\hline 
$\omega_{1}$ & 1313 & 1305 (51) & 1307 (36) & 1345\tabularnewline
\hline 
$\omega_{2}$ & 1535 & 1529 (48) & 1530 (32) & 1570\tabularnewline
\hline 
2$\omega_{1}$ & 2624 & 2594 (80) & 2614 (50) & 2690\tabularnewline
\hline 
$\omega_{1}+\omega_{2}$ & 2836 & 2820 (61) & 2839 (36) & 2915\tabularnewline
\hline 
$\omega_{3}$ & 2949 & 2948 (N/A) & 2950 (34) & 3036\tabularnewline
\hline 
$\omega_{4}$ & 3053 & 3050 (58) & 3058 (38) & 3157\tabularnewline
\hline 
ZPE & 9707 & 9696 (46) & 9688 (34) & 9842\tabularnewline
\hline 
MAE & - & 11 & 7 & 77\tabularnewline
\hline 
\end{tabular}
\end{table}
The investigated methane overtone (level 1.1) and combination excitation
(level 1.1 2.1) are basically harmonic at the quantum mechanical level.
This has permitted to get excellent AS-SCIVR estimates also for them
by means of a single simulation started from harmonic ZPE quantization. 

\newpage

\section{Summary and Conclusions\label{sec:Summary-and-Conclusions}}

We have introduced a new strategy, AS SCIVR, to perform quantum vibrational
simulations. It is made of a preliminary adiabatic switching procedure
for initial conditions followed by a semiclassical spectroscopic calculation.
The two main advances introduced by the new technique lie in the very
limited number of numerically unstable semiclassical trajectories
and the reduced width of spectroscopic signals. Accuracy, which was
actually already very good for TA-SCIVR simulations, is also improved,
especially when the AS evolution is initiated from the appropriate
harmonic quantization. In our AS-SCIVR simulations the mean absolute
error with respect to quantum calculations was below 10 cm\textsuperscript{-1}.
Furthermore, an AS-SCIVR simulation started from the harmonic ZPE
quantization is able to return very good estimates for fundamental
frequencies, while it gives a less accurate representation of overtones
as it provides merely harmonic values. In these aspects AS SCIVR resembles
the MC-SCIVR approach.

While discussing results for H\textsubscript{2}O we noticed that
it is the ZPE eigenvalue rather than frequency estimates that carries
most of the inaccuracy. This is due to the presence of a small amount
of rotational angular momentum in the AS procedure, since molecules
are prepared out of equilibrium and normal modes are no longer correctly
defined for pure vibrations. We tried to remove the angular momentum
before the adiabatic switching dynamics was started obtaining indeed
a better ZPE value. For instance, the ZPE of water shifted from 4637
to 4654 cm\textsuperscript{-1}, closer to the quantum mechanical
benchmark at 4660 cm\textsuperscript{-1}. However, we found that
spectral signals were irregular in shape and much larger, and the
technique lost one of its peculiar features making the gain in ZPE
accuracy not particularly appealing. Furthermore, frequency values,
i.e. the data of interest for comparison to experiments, are calculated
by difference between two eigenenergies, so the angular momentum effect
cancels out and estimates are accurate. In fact, AS-SCIVR MAE values,
when restricted to fundamentals only, decrease to 3 cm\textsuperscript{-1}
for H\textsubscript{2}O, and 4 cm\textsuperscript{-1}for CH\textsubscript{4}.
An additional confirmation that the angular momentum component is
a possible source of inaccuracy in estimating SC eigenvalues comes
from our application to the Henon-Heiles model potential. In that
case the system was defined in normal modes with no rotation allowed.
Remarkably, we were able to reproduce a set of 8 eigenenergies exactly.
As for the precision of results, it is known that the presence of
rotational angular momentum may affect the width of SC signals. Nagy
and Lendvay's internal coordinate adiabatic switching is angular-momentum
free and could be helpful, but our approach, which interfaces straightforwardly
with the SC calculations, brings in most of the advance overperforming
TA SCIVR neatly and providing very accurate and precise results.

Another important feature of AS SCIVR is that it can be readily interfaced
with any pre-existing semiclassical approach including MC SCIVR and
DC SCIVR at the affordable cost (with respect to Hessian matrix calculations)
cost of just an additional dynamics. This opens up the possibility
to achieve a better resolution in simulations involving large dimensional
systems, which may help enormously in the difficult assignment of
the crowded regions of the spectrum. Furthermore, the diminished probability
of trajectory rejection is encouraging, since it increases the probability
that in \emph{ab initio} simulations based on a single trajectory
the standard prefactor is adopted for the entire dynamics without
introduction of any approximation. In fact, rejection is virtually
absent (rejection percentage < 5\%) in all our AS-SCIVR calculations,
with the exception of H\textsubscript{2}CO. Even in this case, though,
only at the larger time studied and adopting a very tight threshold,
the AS-SCIVR procedure appears to be less effective. However, these
conditions are way too stringent for our \emph{ab initio} on-the-fly
simulations, for which we generally employ a threshold $\sigma$ =
10\textsuperscript{-2} and a dynamics about 0.6 ps long.

Finally, in addition to improve SC simulations of large dimensional
systems, the AS-SCIVR method we have benchmarked in this paper might
also serve in perspective as an innovative tool for the semiclassical
investigation of floppy systems, which constitute very complex research
topics on their own. The initial setup would require a particular
care in defining normal modes and the conversion matrix between them
and Cartesian coordinates, as largely debated in Ref. \citenum{Bertaina_Ceotto_Zundel_2019}.
Currently the semiclassical study of these systems needs adoption
of particular devices mainly consisting in the removal of energy from
the large amplitude, low frequency modes.\citep{Ceotto_watercluster_18,Bertaina_Ceotto_Zundel_2019}
AS SCIVR may help avoid this artefact yielding more accurate and precise
frequency estimates. 
\begin{acknowledgments}
Authors acknowledge financial support from the European Research Council
(Grant Agreement No. (647107)\textemdash SEMICOMPLEX\textemdash ERC-
2014-CoG) under the European Union\textquoteright s Horizon 2020 research
and innovation programme, and from the Italian Ministery of Education,
University, and Research (MIUR) (FARE programme R16KN7XBRB- project
QURE). Part of the cpu time was provided by CINECA (Italian Supercomputing
Center) under ISCRAB project ``QUASP''.
\end{acknowledgments}

\bibliographystyle{aipnum4-1}
\bibliography{Biblio_AUG2019}

\end{document}